\documentclass[conference]{IEEEtran}

\IEEEoverridecommandlockouts

\usepackage{cite}

\usepackage{amsmath,amssymb,amsfonts}

\usepackage{algorithmic}

\usepackage{graphicx}

\usepackage{textcomp}

\usepackage{xcolor}

\usepackage{multirow}

\def\BibTeX{{\rm B\kern-.05em{\sc i\kern-.025em b}\kern-.08em

    T\kern-.1667em\lower.7ex\hbox{E}\kern-.125emX}}

\begin{document}

\title{Identifying Distribution Network Faults Using Adaptive Transition Probability}

% \\




% }

\author{\IEEEauthorblockN{Xinliang Ma}
\IEEEauthorblockA{\textit{Luohe Power Supply Company}\\
Luohe, China}
\and
\IEEEauthorblockN{Weihua Liu}
\IEEEauthorblockA{\textit{Luohe Power Supply Company}\\
Luohe, China}
\and
\IEEEauthorblockN{Bingying Jin}
\IEEEauthorblockA{\textit{Academy of Electrical Engineering} \\
\textit{Shanghai Jiaotong University}\\
Shanghai, China}
}

\maketitle

\begin{abstract}
A novel approach is suggested for improving the accuracy of fault detection in distribution networks. This technique combines adaptive probability learning and waveform decomposition to optimize the similarity of features. Its objective is to discover the most appropriate linear mapping between simulated and real data to minimize distribution differences. By aligning the data in the same feature space, the proposed method effectively overcomes the challenge posed by limited sample size when identifying faults and classifying real data in distribution networks. Experimental results utilizing simulated system data and real field data demonstrate that this approach outperforms commonly used classification models such as convolutional neural networks, support vector machines, and k-nearest neighbors, especially under adaptive learning conditions. Consequently, this research provides a fresh perspective on fault detection in distribution networks, particularly when adaptive learning conditions are employed.
\end{abstract}


\section{Introduction}

Distribution networks play a critical role in the electricity supply system by serving end-users and ensuring power quality, operational efficiency, and innovative customer services \cite{xiao2020overview} \cite{sun2020development}. Despite their extensive coverage, varied equipment, and relatively low replacement costs, distribution networks are often neglected by power utilities when it comes to ensuring reliable supply \cite{wang2020review}. Consequently, the development of fault identification technologies in these networks has been slow \cite{wang2020review}. Currently, most faults in distribution networks are only addressed through repairs after they occur, leading to significant service disruptions for users. However, as the power grid evolves and expectations for supply reliability increase, power utilities are now placing more importance on predicting and diagnosing equipment faults in distribution networks. This shift aims to promptly address safety risks and minimize the frequency of power outages.

Faults in equipment within distribution networks can occur suddenly or gradually over time \cite{xiong2020detection}. One common fault is a ground fault, where protective devices isolate the faulty portion and restore normal operation once the fault is resolved \cite{yang2020location}. However, the electrical arcing during the fault can cause irreparable damage to the insulation. If this process repeats multiple times, it can lead to insulation degradation and eventual breakdown \cite{yang2019simulation}. These initial phase faults, referred to as "early-stage faults" in this article, are often disregarded by power utilities but contain valuable information about the insulation \cite{fang2020development}. If effectively utilized, the waveforms associated with these faults can provide early warnings of faults in distribution networks and improve supply reliability \cite{li2020anomaly} \cite{chen2020research}.

Due to the complex nature of distribution network structures, traditional waveform analysis methods based on mechanisms are not efficient \cite{zheng2021rsspn}. However, the integration of multiple sensors has allowed for the adoption of data-driven models in this field. The identification of faults in distribution networks is a challenging task, particularly when it involves electrical arcing in ground faults, which are characterized by randomness and uncertainty \cite{liu2022high}. Obtaining sufficient data for training models is difficult due to the rarity of self-recoverable faults \cite{tao2020parallel}. As a result, many fault identification algorithms rely on simulated or experimental data for developing and testing models. Only a small number of algorithms incorporate real field data, which is more complex and influenced by multiple interfering factors \cite{ju2020study} \cite{xu2020detection}. Therefore, it is crucial to assess model performance using real field data \cite{zhang2019fault}\cite{liang2020novel}. Furthermore, modern artificial intelligence techniques such as diverse convolutional neural networks often lack interpretability. The features extracted from these models are not easily understandable to humans, making it challenging to evaluate the quality of features and incorporate prior knowledge.

We present a proposed solution to tackle the previously mentioned challenges associated with identifying faults in distribution networks. Our approach incorporates adaptive probability learning, which entails training the model using simulated data and testing it with actual field data. By evaluating feature similarity and extracting universal features, our adaptive probability learning algorithm overcomes the difficulties posed by varying network structures, line parameters, and operating conditions. This is crucial as the distributions of simulated and real data often diverge in these aspects.

Our method comprises two stages: waveform decomposition to obtain feature vectors in the first stage, and linear mapping in the second stage for dimension reduction and feature reconstruction. We determine the optimal linear mapping by maximizing the likelihood of consistent reconstruction. Additionally, we utilize clustering in the reduced-feature space to classify events. Compared to other approaches, our model effectively addresses the disparities between simulated and real data and offers strong interpretability. This presents a fresh and efficient strategy for fault identification in distribution networks.

\section{Adaptive Probability Learning}

An objective of adaptive learning is to address the discrepancy in data distributions between simulated and real-world data. One proposed strategy to overcome this problem involves measuring a model's performance across different data contexts using feature similarity as a metric \cite{liang2020novel}\cite{liu2019prediction}. On the other hand, adaptive probability learning utilizes probability to evaluate the similarity between different features by calculating the reconstruction error, providing valuable insights \cite{haeusser2017associative}\cite{xiong2020incipient}\cite{wu2020harmonic}.

\subsection{Adaptive Learning}

Adaptive learning relies on two kinds of data: simulation data and real-world data \cite{wang2020power}. Simulation data ($D_{s}$) includes waveforms and event classes from simulated events, while real-world data ($D_{t}$) consists of waveforms of events with unknown classes. The main idea behind adaptive learning is that the event categories in the simulation data are similar to those in the real-world data. This similarity allows us to make informed guesses about the event categories in the real-world data by utilizing the knowledge gained from the simulated data \cite{liu2022method}. However, it is important to note that these two datasets often have different distributions due to variations in grid structure, line parameters, and operational conditions. In simpler terms, the probability distribution of the simulation data ($P_{s}\left(x_{i}^{s}, y_{i}^{s}\right)$) may not be the same as that of the real-world data ($P_{t}\left(x_{i}^{t}, y_{i}^{t}\right)$). Therefore, adapting knowledge from one domain to another presents challenges and complexities.

\begin{figure}[htbp]
\centerline{\includegraphics[width=0.6\linewidth]{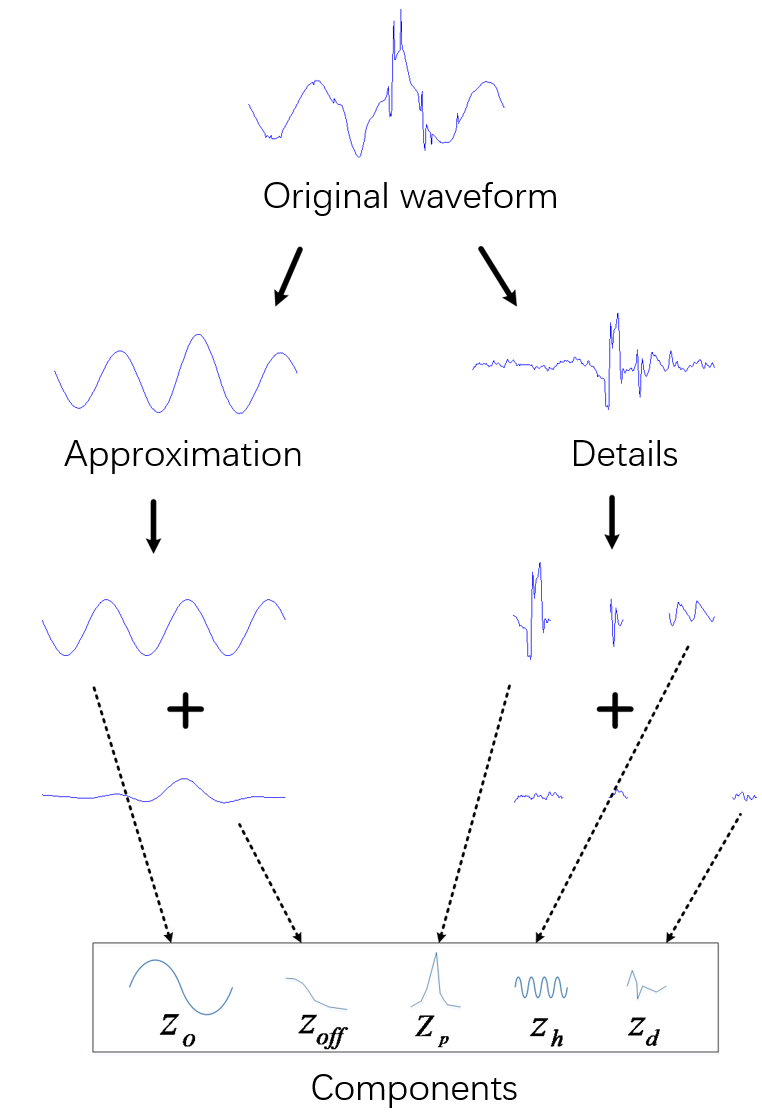}}
\caption{Illustration of Waveform Decomposition}
\label{fig}
\end{figure}

It is important to consider that the performance of a classification model, as measured by the error $L_s$ on simulated data, may not accurately reflect its performance on real-world data, as indicated by the error $L_t$. This discrepancy arises because we need to determine if the features generated by the model have meaning and can be transferred between simulated and actual data. To address this concern, our proposed approach utilizes waveform decomposition to enhance feature congruence and identify the optimal linear mapping. This relationship can be expressed as $L_t = L_s + L_{sim}$, illustrating that a reliable classification model should not only perform well on simulated data but should also exhibit a high degree of similarity in the extracted features from simulated and actual datasets.

\subsection{Feature Extraction}

The research utilizes a wavelet decomposition-based approach for feature extraction, which is advantageous compared to deep neural networks. One advantage is that it necessitates less data, making it appropriate for situations with limited samples such as distribution network fault diagnosis \cite{li2020fault}. Another advantage is that the extracted features are highly interpretable, facilitating the incorporation of prior knowledge and enhancing accuracy \cite{xie2016development}.

\begin{figure}[htbp]
\centerline{\includegraphics[width=0.8\linewidth]{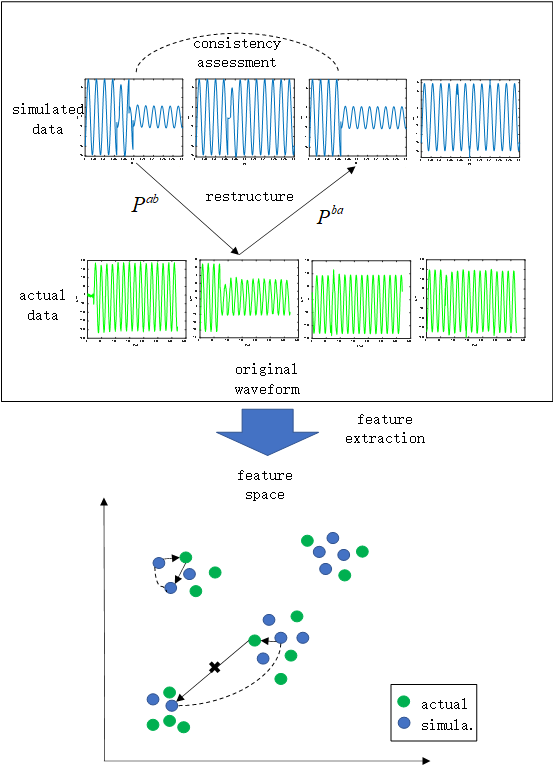}}
\caption{An illustration of principles of reconstruction error}
\label{fig}
\end{figure}

The method steps are as follows: first, waveforms are decomposed into approximate and detail components through wavelet transformation. The approximate component reflects the overall shape of the waveform, while the detail component reflects distortions. Based on this decomposition, the fundamental component $Z_{o}$ and bias $Z_{\text {off }}$ are extracted from the approximate component, and pulse $z_{p}$, harmonic $Z_{h}$, and distortion $Z_{d}$ are extracted from the detail component. For different components, corresponding features are extracted. For example, the fundamental component $Z_{o}$ corresponds to features like amplitude $A_{o}$ and frequency $f_{o}$, bias $z_{\text {off }}$ corresponds to amplitude $A_{\text {off }}$, pulse $z_{p}$ corresponds to peak value $A_{p}$ and pulse width $t_{p}$, harmonic $Z_{h}$ corresponds to amplitude $A_{h}$ and frequency $f_{h}$, and distortion $z_{d}$ corresponds to distortion factor $w_{d}$. All features are normalized to eliminate scale effects. In addition to the features of the components themselves, the time intervals between components $t\left(z_{i}, z_{i+1}\right)$ are also considered, where $z_{i}$ represents the $i$-th component. Figure 1 gives an illustration of waveform decomposition, and after this decomposition, any waveform $w=\left\{I_{\mathrm{A}}, I_{\mathrm{B}}, I_{\mathrm{C}}, U_{\mathrm{A}}, U_{\mathrm{B}}, U_{C}\right\}$ can be uniquely determined by a feature vector $\phi(w)=\left[A_{o}, f_{o}, A_{o f f}, A_{p}, t_{p}, A_{h}, f_{h}, w_{d}, t\left(z_{i}, z_{i+1}\right)\right]$.

\subsection{Adaptive Probabilistic Learning}

The problem involves analyzing both simulated data ($x_{i}^{s}$) and real data ($x_{j}^{t}$). We know the category of the simulated data ($y_{i}^{s}$), but the category of the real data ($y_{j}^{t}$) is unknown. We extract features from the data, resulting in feature vectors denoted as $A_{i}:=\phi\left(x_{i}^{s}\right)$ and $B_{j}:=\phi\left(x_{j}^{t}\right)$. Since the feature vectors $A_{i}$ and $B_{j}$ have high dimensionality, we use a linear mapping $\varphi$ to reduce their dimensions, resulting in reduced feature vectors $A_{i}^{\prime}:=\varphi\left(A_{i}\right)$ and $B_{j}^{\prime}:=\varphi\left(B_{j}\right)$. We then assess the similarity between $A_{i}^{\prime}$ and $B_{k}^{\prime}$ by calculating the reconstruction error. This allows us to determine the probability of transforming between $A_{i}^{\prime}$ and $B_{k}^{\prime}$ and vice versa. It is important to note that although $A_{i}^{\prime}$ and $A_{j}^{\prime}$ may be different, their respective categories ($y_{i}^{s}$ and $y_{j}^{s}$) must be the same. Figure 2 illustrates the concept of the reconstruction error.

First, to determine the likelihood of transforming $A_{i}^{\prime}$ into $B_{k}^{\prime}$, we compute the inner product of the two vectors, denoted as $M_{ik}=\left\langle A_{i}^{\prime}, B_{k}^{\prime}\right\rangle$. Next, we evaluate the correlation probability between the two vectors using a undisclosed formula.

$$
P_{i k}^{a b}=P(B_{k}'|A_{i}'):=\frac{\exp(M_{i k})}{\sum_{k'} \exp(M_{i k'})}
$$

Likewise, the likelihood of $B_{k}'$ changing into $A_{i}'$ can be represented as follows:

$$
P_{k j}^{b a}=P(A_{j}'|B_{k}'):=\frac{\exp(M_{k j})}{\sum_{k'} \exp(M_{k' j})}
$$

Therefore, converting from $A_{i}^{\prime}$ to $B_{k}^{\prime}$ and then to $A_{j}^{\prime}$ in this cycle has a probability of:

$$
P_{i j}^{a b a}=\left(P^{a b} P^{b a}\right)_{i j}=\sum_{k} P_{i k}^{a b} P_{k j}^{b a}
$$

If the values $y_{i}^{s}$ and $y_{j}^{s}$ are in agreement, then the anticipated distribution of $P_{i j}^{a b a}$ can be expressed as:

$$
T_{i j}^{a b a}=\frac{1}{2}\left(P_{i j}^{a b a}+P_{i j}^{a b a}\right)
$$

$$
T_{ij}:= \begin{cases}1 / N_{c}(y_{i}^{s}) & A_{i}^{\prime}, A_{j}^{\prime} \text { have the same category } \\ 0 & \text { otherwise }\end{cases}
$$

The measure $H$ quantifies the difference between $P_{ij}^{aba}$ and $T$. In order to calculate $P_{ij}^{aba}$, we use the count of instances that belong to category $y_{i}^{s}$ in dataset $D_{s}$, denoted as $N_{c}(y_{i}^{s})$.

$$
L_{w}:=H(P_{ij}^{aba}, T)
$$

To enhance the integration of real data in the reconstruction process, we consider traversal error. This error evaluates the likelihood of each genuine data element being included in the reconstruction procedure.

$$
L_{v}=H(P^{visit}, V)
$$

$$
P_{k}^{v}=\sum_{x_{i}^{s}} P_{i k}^{ab}, V_{k}=1/|D_{t}|
$$

During the training process of the model, the overall error, represented as $L_{sim}$, is calculated by combining the errors that arise from measuring feature similarity and analyzing simulation data. These calculations entail evaluating and integrating errors from both categories.

\section{Experimental Data and Procedure}

In order to verify and assess the reliability of the proposed model, a series of experiments are conducted, utilizing both simulated data and real-world data gathered in real-life situations. The key objective of adaptive learning is to initially train the model using simulated data and subsequently enable it to autonomously identify and categorize real data.

\subsection{Simulation system}

\begin{figure}[htbp]
\centerline{\includegraphics[width=0.7\linewidth]{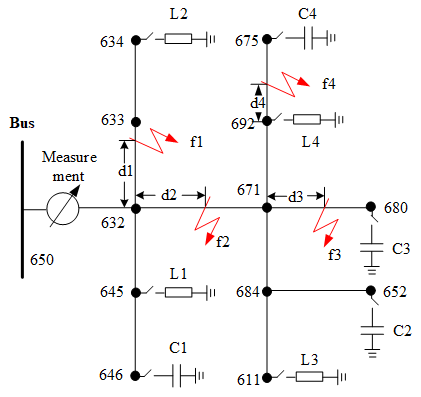}}
\caption{Configuration of the simulation system}
\label{fig}
\end{figure}

The simulation system shown in Figure 3 is based on the IEEE 13 node model and operates at a voltage level of 10 kV and frequency of 50 Hz. A sampling frequency of 4 kHz is used in the simulation. The figure indicates the fault location and load conditions. The simulation system is created using PSCAD software \cite{liu2021method}. To simulate faults, the Kizilcay arc model is employed, which captures the dynamic behavior of the arc using control theory principles, specifically focusing on the energy balance within the arc column. The mathematical expression for this arc model is provided as follows: 

$$
\begin{gathered}
\frac{d g(t)}{d t}=\frac{1}{\tau}\left(\frac{\left|i_{f}(t)\right|}{u_{o}+r_{o}\left|i_{f}(t)\right|}-g(t)\right) \\
g(t)=\frac{u_{f}(t)}{i_{f}(t)}
\end{gathered}
$$

Different variables such as fault impedance, fault starting angle, fault distance, line parameters, load parameters, and noise levels were modified to simulate network fault data under different conditions. These modifications allowed for the creation of simulation data for four types of events: single-frequency early fault, multi-frequency early fault, permanent fault, and transient interference \cite{liu2020basic}. Figure 3 indicates the potential locations of early faults, permanent faults, capacitors, and loads. A total of 10 sets of simulation data were randomly generated for all possible scenarios. The arc conductance $g(t)$, arc current $i_f(t)$, and arc voltage $u_f(t)$ are measured in S/m, A, and V, respectively. The arc time constant $\tau$, arc characteristic resistance $r_o$, and arc characteristic voltage $u_o$ are measured in seconds, ohms, and volts, respectively. The parameter ranges for $\tau$, $u_o$, and $r_o$ are 0.2-0.4 ms, 300-4000 V, and 0.01-0.015 $\Omega$, respectively.

\begin{figure}[htbp]
\centerline{\includegraphics[width=0.8\linewidth]{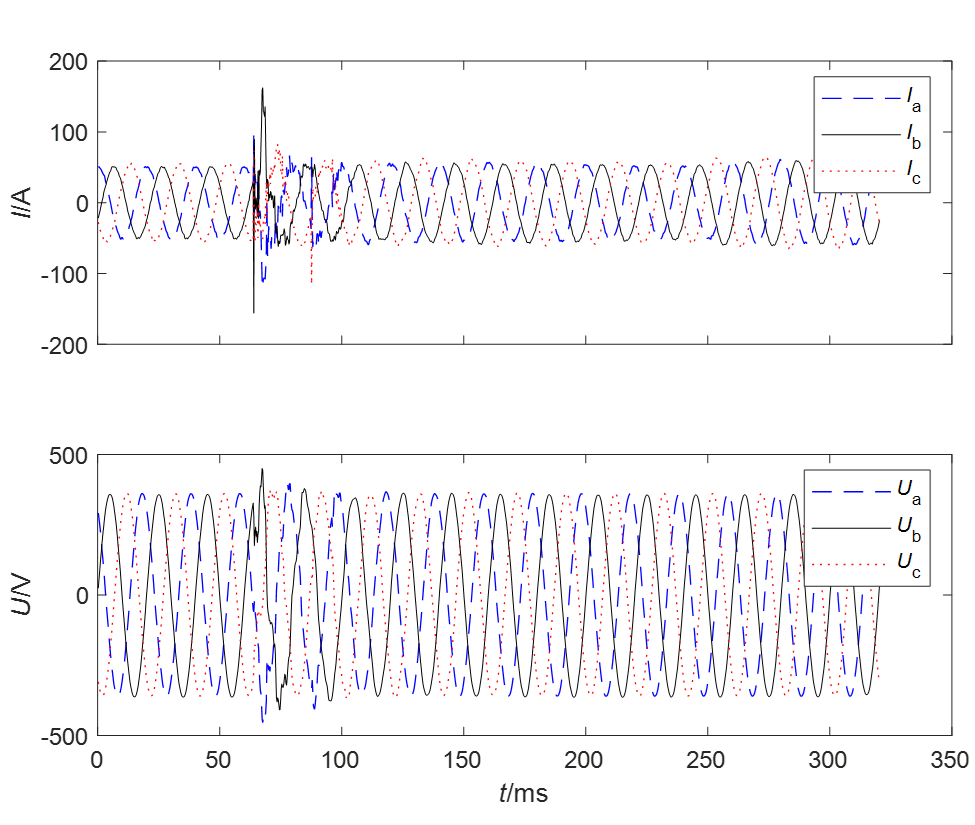}}
\end{figure}

\begin{figure}[htbp]
\centerline{\includegraphics[width=0.8\linewidth]{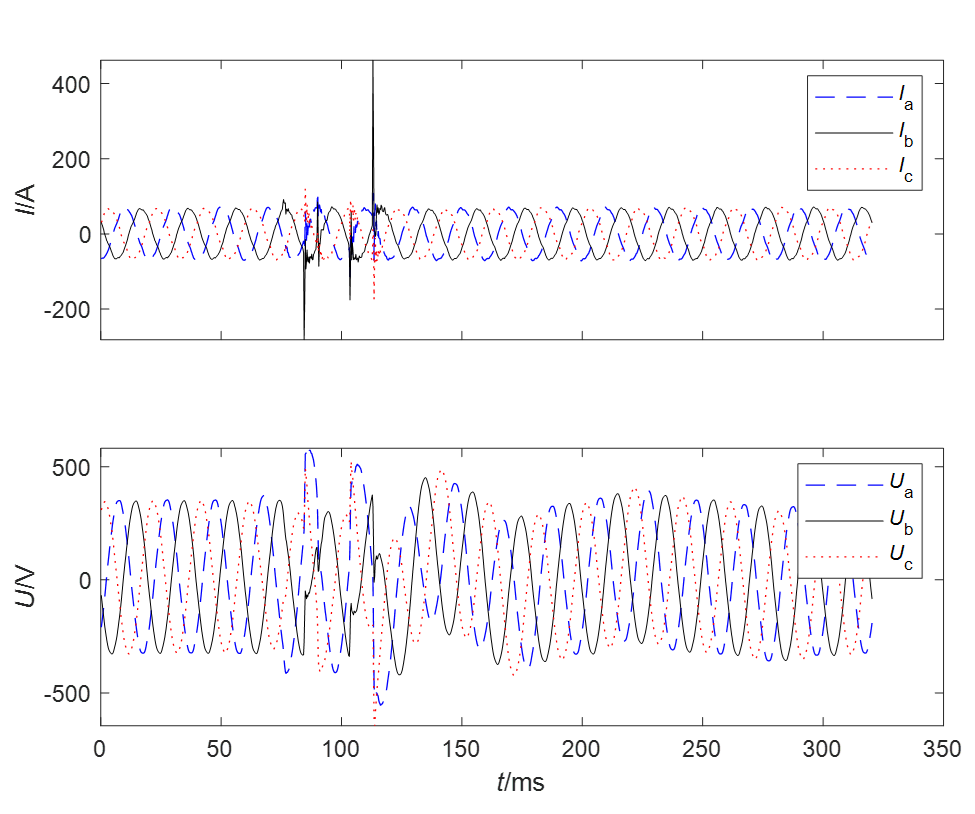}}
\end{figure}

\begin{figure}[htbp]
\centerline{\includegraphics[width=0.8\linewidth]{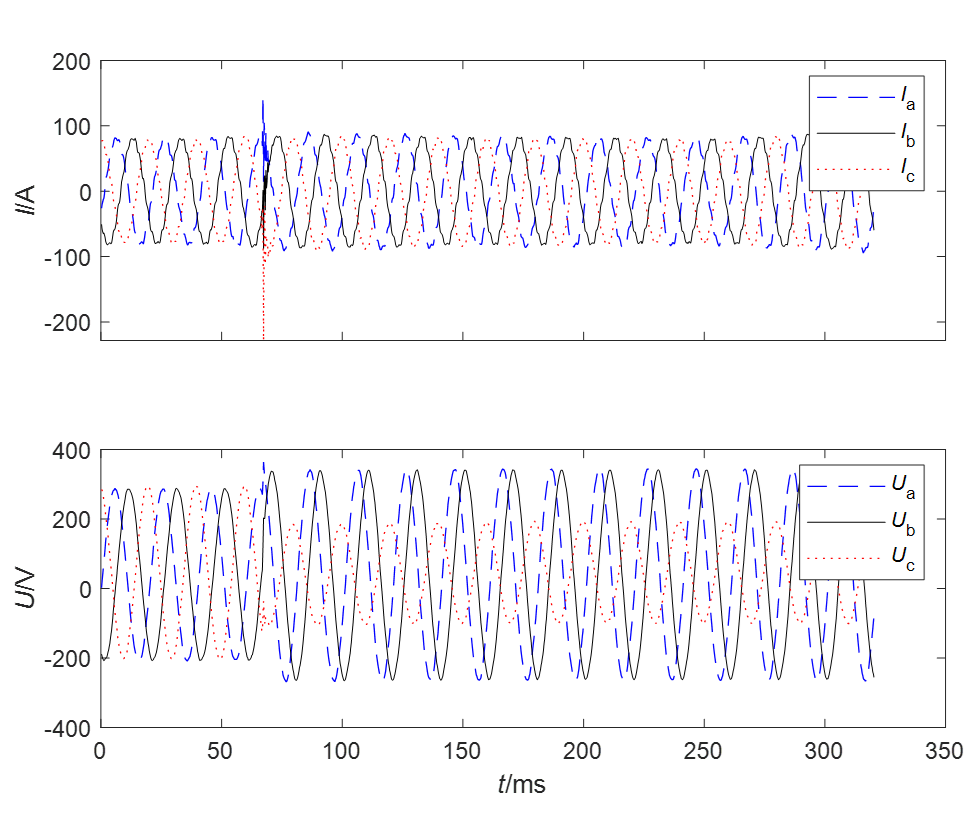}}
\end{figure}

\begin{figure}[htbp]
\centerline{\includegraphics[width=0.8\linewidth]{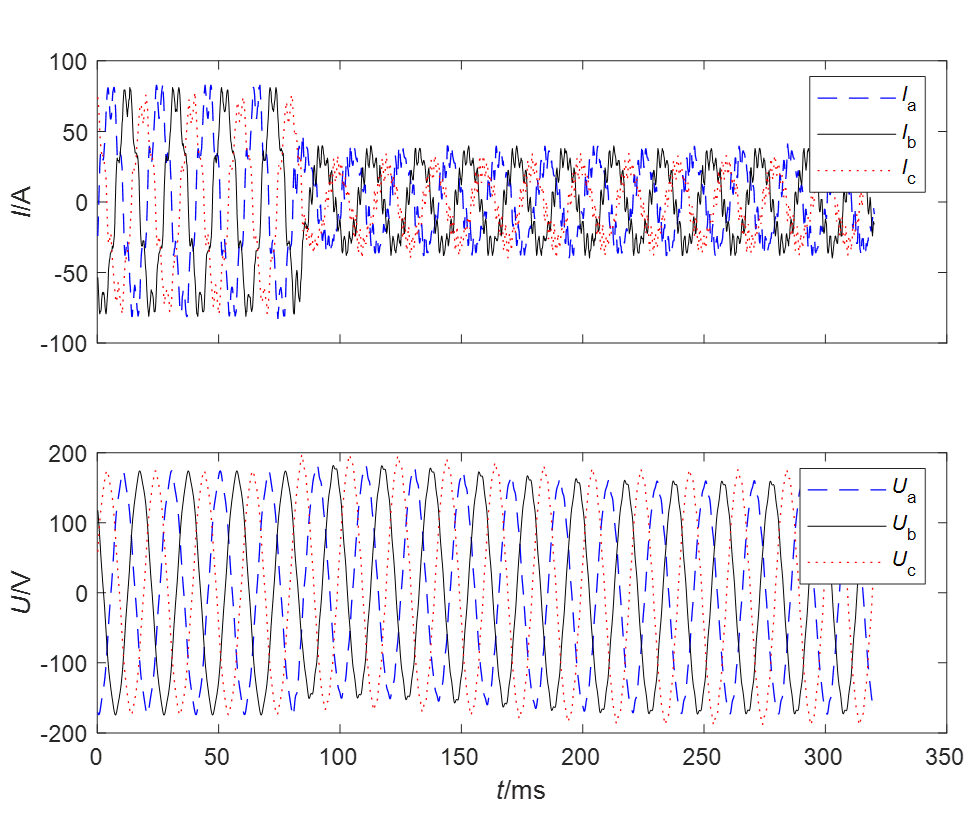}}
\caption{Typical waveform for different types of events}
\label{fig}
\end{figure}

\subsection{Actual Data}

Between February and May 2021, data was gathered in Guangdong Province from fault detection devices installed on a 10 kV overhead line. The system utilizes low-current grounding and the fault detection device samples voltage and current signals at a frequency of 4096 Hz. Recording commences when the voltage or current signal surpasses a predefined threshold. The device captures the three-phase voltage and current waveforms before and after a fault occurrence, with a recording duration of sixteen cycles.

\begin{table}[htbp]

\caption{Distribution of event type}

\begin{center}

\begin{tabular}{lcccc}

\hline

Data Type & SIF & MIF & PF & TD \\

\hline

Simulation Data & 80 & 80 & 80 & 80 \\

Actual Data & 71 & 64 & 93 & 88 \\

Total & 151 & 144 & 173 & 168 \\

\hline

\end{tabular}

\end{center}

\end{table}

Events in this classification are categorized based on waveform analysis and onsite confirmation of the fault's cause. There are three types of events: Early faults, Permanent faults, and Transient interferences. Early faults are transient faults that can be recovered and are further divided into single-cycle early faults and multi-cycle early faults, indicating varying levels of severity. Permanent faults cannot be self-recovered and require intervention from protective devices \cite{xie2016development}. While fault detection devices can also be triggered by overvoltages caused by operations and lightning, these overvoltages are not considered faults but are categorized as transient interferences \cite{zhang2020research}. Figure 4 illustrates typical waveforms associated with each event type. In addition, Table 1 provides statistical data on the quantities of each event type.

\subsection{Experimental Procedure}

Two experiments were carried out to assess the adaptability of the model in this study. In the first experiment, the model was trained using all simulated data, while a random sample of actual data was used for validation. The remaining actual data was then used as the test set, with known labels obtained from the validation set. The second experiment followed a similar setup, with all simulated data used for training, a random sample of actual data used for validation, and the remaining actual data used for testing. However, in this case, the labels of the validation set were unknown. It is important to note that each of these three sets serves a specific purpose: the training set is used to train the model, the validation set assesses performance and tunes hyperparameters, and the test set evaluates the final model's performance.

The experimental data is divided into three groups \cite{zhang2019power}: the training set comprising 320 samples, the validation set consisting of 160 samples, and the test set containing 156 samples. To minimize the impact of event type distribution, each experiment is replicated 10 times, and the average performance across these 10 iterations is measured. The F1 score is utilized as the evaluation metric to assess the model's performance.

\section{Experimental Results and Analysis}

In this section, we will compare the adaptive probabilistic learning model introduced in this study with three commonly employed classifiers: Convolutional Neural Network (CNN), Support Vector Machine (SVM), and K-Nearest Neighbors (KNN) algorithm. The objective is to illustrate the superiority of our proposed method. Unlike traditional classifiers, our adaptive probabilistic learning model considers the variances between simulated and real data distributions and integrates the notion of feature similarity to tackle this problem. As a result, it achieves substantially improved performance.

\subsection{Adaptive Probabilistic Learning}

The adaptive probabilistic concept learning model undergoes a training process that involves breaking down waveforms into different components, such as fundamental wave, bias, pulse, harmonic, and distortion. For each component, feature values and time intervals are calculated, and these feature vectors are then reduced in dimensionality through a linear mapping step. The similarity of features is measured by the reconstruction error. The training of the model involves estimating the actual data error, which combines the feature similarity error and the classification error of simulated data. This estimation helps determine the optimal parameters of the linear mapping. During testing, test waveforms are decomposed and mapped onto the feature space using linear mapping. Once in this space, they are clustered to make predictions about the corresponding event types. In Experiment 1, the validation set contains the actual data labels. This allows the model trained using the aforementioned method to directly predict the validation set. Using these predicted results, the best model can be identified by comparing them with the true labels. Subsequently, this best model is applied to the test set to generate the final test results.

The validation set used in Experiment 2 includes unlabeled data. As a result, the model can be directly used on the test set to obtain the final test results.

Table 2 presents evidence of how the adaptive probabilistic learning model effectively handles diverse faults. The model achieves this by taking into account the dissimilarities in distribution between simulated and real data and utilizing the similarities in characteristics to establish a correlation between the model's errors on both types of data. The central concept involves identifying a suitable conversion technique that transforms the original waveform into a feature vector space. This enables precise categorization of both simulated and real data within this space, ensuring that similar data is clustered together while dissimilar data is dispersed.

\begin{table}[htbp]

\caption{F1 score of different models}

\begin{center}

\begin{tabular}{ccccccc}

\hline

Model & Exp. & SIF & MIF & PF & TD & Ave. \\

\hline

APL & 1  & 0.910 & 0.945 & 0.972 & 0.897 & 0.931 \\

 & 2 & 0.874 & 0.904 & 0.951 & 0.847 & 0.894 \\

CNN & 1 & 0.742 & 0.784 & 0.801 & 0.714 & 0.760 \\

& 2 & 0.701 & 0.722 & 0.741 & 0.684 & 0.712 \\

SVM & 1 & 0.804 & 0.791 & 0.831 & 0.763 & 0.797 \\

& 2 & 0.714	& 0.721	& 0.730	& 0.691	& 0.714 \\

KNN & 1 & 0.725	&0.740	&0.734&	0.667&	0.717\\

& 2 & 0.684	&0.707	&0.700	&0.624&	0.679\\
\hline
\end{tabular}

\end{center}

\end{table}

\subsection{Comparing with other models}

The experiment utilized a convolutional neural network model based on the architecture of AlexNet. The model consisted of 5 convolutional layers and 3 fully connected layers. The input layer had dimensions of 1 x N x 6, where N represented the length of the sample including 6 groups of waveforms. The first convolutional layer had a kernel size of 1 x 41, a stride of 20, and 40 convolutional kernels. The second convolutional layer had a kernel size of 1 x 20, a stride of 10, and 20 convolutional kernels. The third, fourth, and fifth convolutional layers had kernel sizes of 1 x 10, strides of 5, and 10 convolutional kernels each. The three fully connected layers had sizes of 512, 512, and 4, respectively, with the last layer representing the number of output classes. The process involved extracting and linearly transforming the input waveform, with the final output indicating the probability of the event belonging to each class. Choosing the appropriate kernel function was crucial for the support vector machine model, and in Experiment 2, the kernel function category with the highest accuracy in the training set was selected. Both Experiment 1 and 2 confirmed the selection of the polynomial kernel function type.

In the K-nearest neighbor algorithm, the selection of the hyperparameter K involves testing different values and comparing their classification accuracy on the validation set. The value of K that yields the highest accuracy on the training set is chosen since the labels of the validation set are unknown in Experiment 2. The optimal K value was found to be 5 in Experiment 1 and 10 in Experiment 2. Table 2 presents the F1 scores of different models, with the adaptive probability learning model showing significantly higher classification accuracy compared to the other three models. This is because the model takes into account the dissimilarities between the training and test sets, and the extracted features exhibit high similarity between the two sets, indicating that the model captures general features well. On the other hand, the other three models perform well on the training set but poorly on the test set due to differences in data distribution. Additionally, Experiment 1 demonstrates significantly higher accuracy than Experiment 2, suggesting that the partially known labels in Experiment 1 help the model overcome distribution differences and achieve better classification accuracy. The difference in accuracy between Experiment 1 and Experiment 2 highlights the adaptive learning capability of the model. It is worth noting that the adaptive probability learning model outperforms the other three models in terms of this ability, indicating reduced reliance on actual data labels.

\begin{figure}[htbp]
\centerline{\includegraphics[width=0.9\linewidth]{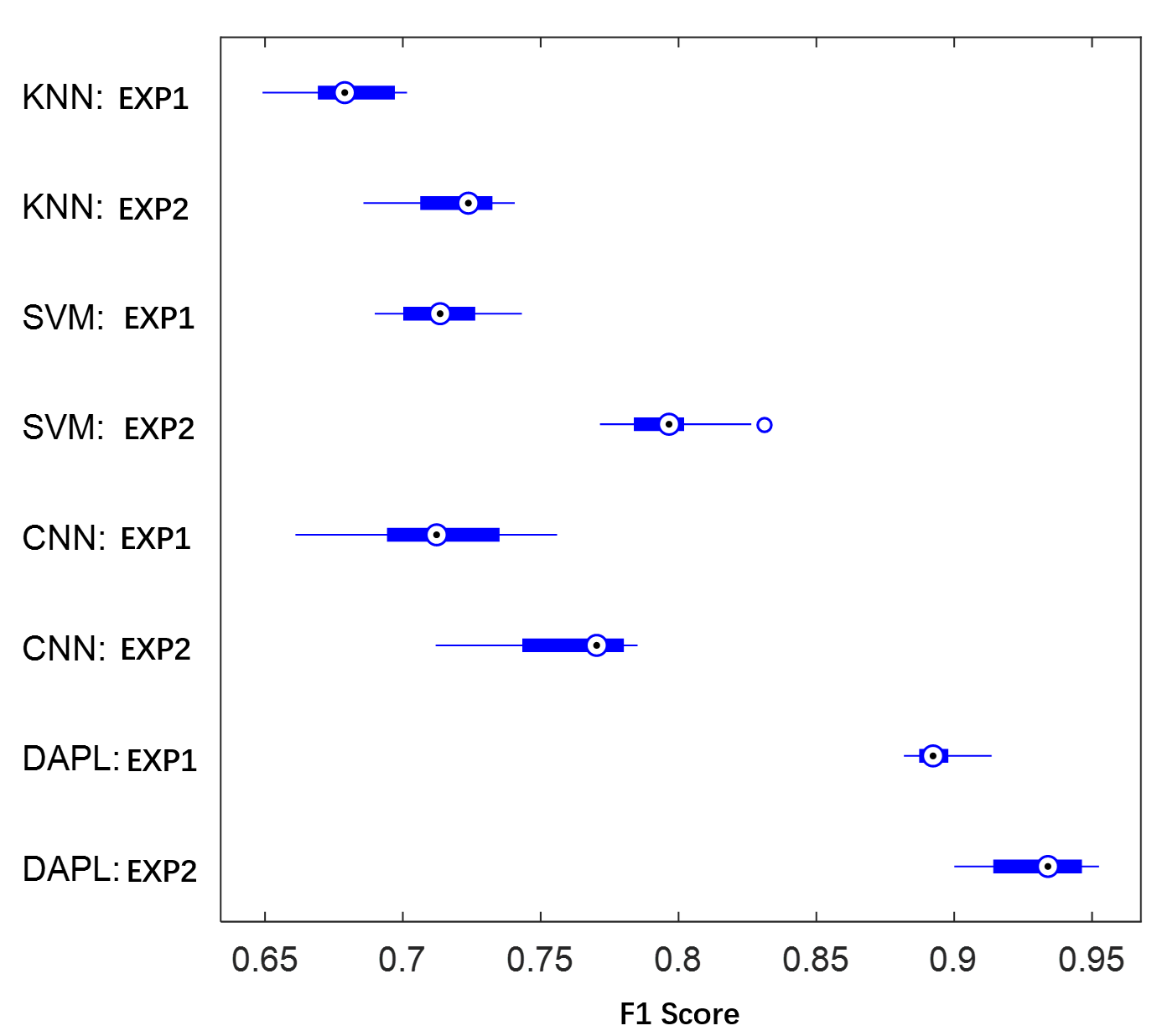}}
\caption{Distribution of Average F1 Score over classes for different models}
\label{fig}
\end{figure}

Figure 5 illustrates the stability of the average class F1 scores \cite{krizhevsky2012imagenet} for the four models, indicating whether their classification accuracy varies with changes in event distribution. The adaptive probability learning model demonstrates highly consistent classification accuracy. In experiment 1, its accuracy consistently falls within the range of 0.90 to 0.95, while in experiment 2, it remains concentrated around 0.90. In contrast, the convolutional neural network exhibits more scattered accuracy, suggesting a greater vulnerability to variations in data type distribution. This can be attributed to its numerous parameters and reliance on substantial amounts of data for determining network weights. Consequently, in scenarios with limited samples and significant variations in event types, training results often differ, leading to fluctuating accuracy. While the support vector machine and K-nearest neighbor algorithm also display relatively stable classification accuracy, their overall performance level is low and not suitable for practical real-world scenarios. In conclusion, the proposed method is best suited for identifying faults in distribution networks.

\section{Conclusion}

The scarcity of training samples is a major hurdle when it comes to identifying faults in distribution networks. To address this problem, the adaptive probability learning method utilizes a two-step approach. First, it extracts feature vectors by breaking down waveforms, and then it reduces the dimensionality through linear mapping. The model then solves an optimization problem to establish a relationship between errors in simulated and real data, with the goal of maximizing consistency probability during the reconstruction process. This ultimately results in the achievement of an optimal linear mapping.

There are several advantages to using the adaptive probability learning method instead of other methods. One advantage is that it produces extracted features that are easy to interpret, which makes it easier to incorporate prior knowledge. Additionally, this method is able to make good use of simulated data during training, which helps overcome the issue of having limited samples when identifying faults in distribution networks. Finally, by incorporating field actual data, the model's performance can be further improved, allowing maintenance personnel to create a sample library right from the start.

\bibliographystyle{IEEEtran}

\bibliography{ieee_conference}

\begin{thebibliography}{10}
\providecommand{\url}[1]{#1}
\csname url@samestyle\endcsname
\providecommand{\newblock}{\relax}
\providecommand{\bibinfo}[2]{#2}
\providecommand{\BIBentrySTDinterwordspacing}{\spaceskip=0pt\relax}
\providecommand{\BIBentryALTinterwordstretchfactor}{4}
\providecommand{\BIBentryALTinterwordspacing}{\spaceskip=\fontdimen2\font plus
\BIBentryALTinterwordstretchfactor\fontdimen3\font minus \fontdimen4\font\relax}
\providecommand{\BIBforeignlanguage}[2]{{%
\expandafter\ifx\csname l@#1\endcsname\relax
\typeout{** WARNING: IEEEtran.bst: No hyphenation pattern has been}%
\typeout{** loaded for the language `#1'. Using the pattern for}%
\typeout{** the default language instead.}%
\else
\language=\csname l@#1\endcsname
\fi
#2}}
\providecommand{\BIBdecl}{\relax}
\BIBdecl

\bibitem{xiao2020overview}
Z.~Xiao, P.~Xin, Z.~Liu, and et~al., ``An overview of planning technology for active distribution network under the situation of ubiquitous power internet of things,'' \emph{Power System Protection and Control}, vol.~48, no.~3, pp. 43--48, 2020.

\bibitem{sun2020development}
K.~Sun, Q.~Zhang, Z.~Zheng, and et~al., ``Development of distribution network in the future from the perspective of energy internet,'' \emph{Zhejiang Electric Power}, vol.~39, no.~1, pp. 5--12, 2020.

\bibitem{wang2020review}
P.~Wang, G.~Feng, Y.~Wei, and et~al., ``Review of grounding fault testing and design of real experiment field for 10 kv distribution network,'' \emph{Power System Protection and Control}, vol.~48, no.~11, pp. 184--193, 2020.

\bibitem{xiong2020detection}
S.~Xiong, Y.~Liu, J.~Fang, Z.~Cong, Y.~Yan, and X.~Jiang, ``Detection method of incipient faults of power distribution lines,'' \emph{High Voltage Engineering}, vol.~46, no.~11, pp. 259--265, 2020.

\bibitem{yang2020location}
F.~Yang, X.~Zang, H.~Qu, and et~al., ``Location and self-healing of single-phase grounding fault based on distribution automation,'' \emph{Journal of Electric Power Science and Technology}, vol.~35, no.~4, pp. 176--181, 2020.

\bibitem{yang2019simulation}
F.~Yang, X.~Jin, Y.~Shen, and et~al., ``Simulation test and characteristic analysis of grounding fault causes of 10 kv overhead distribution network,'' \emph{Distribution \& Utilization}, vol.~36, no.~3, pp. 37--43, 2019.

\bibitem{fang2020development}
J.~Fang, X.~Peng, T.~Liu, and et~al., ``Development trend and application prospects of big data-based condition monitoring of power apparatus,'' \emph{Power System Protection and Control}, vol.~48, no.~23, pp. 182--192, 2020.

\bibitem{li2020anomaly}
Z.~Li, F.~Qian, A.~Liu, and et~al., ``An anomaly tracing and application model of distribution network topology based on multi-dimensional feature fusion,'' \emph{Zhejiang Electric Power}, vol.~39, no.~7, pp. 71--79, 2020.

\bibitem{chen2020research}
W.~Chen and H.~Chang, ``Research and application of fault analysis and judgment based on multi-source data in distribution network,'' \emph{Electric Power Information and Communication Technology}, vol.~18, no.~12, pp. 43--50, 2020.

\bibitem{zheng2021rsspn}
T.~Zheng, Y.~Liu, Y.~Yan, S.~Xiong, T.~Lin, Y.~Chen, Z.~Wang, and X.~Jiang, ``Rsspn: Robust semi-supervised prototypical network for fault root cause classification in power distribution systems,'' \emph{IEEE Transactions on Power Delivery}, vol.~37, no.~4, pp. 3282--3290, 2021.

\bibitem{liu2022high}
Y.~Liu, Y.~Yingjie, S.~Xiong, L.~Pei, Z.~Li, P.~Xu, L.~Su, X.~Fu, and X.~Jiang, ``High-precision identification method and system for substations,'' Oct.~27 2022, uS Patent App. 17/433,994.

\bibitem{tao2020parallel}
P.~Tao, Y.~Zhang, M.~Li, and et~al., ``Parallel clustering analysis for power consumption data based on graph model,'' \emph{Journal of Electric Power Science and Technology}, vol.~35, no.~6, pp. 146--153, 2020.

\bibitem{ju2020study}
Z.~Ju, Y.~Zhu, Z.~Gao, and et~al., ``Study on transient process of short circuit fault in 10 kv distribution network with series capacitor compensation device,'' \emph{Power Capacitor \& Reactive Power Compensation}, vol.~41, no.~2, pp. 7--13, 2020.

\bibitem{xu2020detection}
Z.~Xu, T.~Ji, W.~Deng, and et~al., ``Detection and identification of cable incipient fault based on autoencoder and gru neural network,'' \emph{Guangdong Electric Power}, vol.~33, no.~9, pp. 27--34, 2020.

\bibitem{zhang2019fault}
G.~Zhang, H.~Pu, and K.~Liu, ``Fault line selection method of small current grounding system based on deep learning,'' \emph{Power Generation Technology}, vol.~40, no.~6, pp. 548--554, 2019.

\bibitem{liang2020novel}
M.~Liang, H.~Teng, X.~Li, and et~al., ``A novel fault location algorithm for radial distribution networks considering the unbalanced characteristics,'' \emph{Journal of Electric Power Science and Technology}, vol.~35, no.~2, pp. 61--68, 2020.

\bibitem{liu2019prediction}
Q.~Liu, W.~Yin, W.~Hu, and et~al., ``Prediction of power harmonic monitoring data based on lstm algorithm,'' \emph{Power Capacitor \& Reactive Power Compensation}, vol.~40, no.~5, pp. 139--145, 2019.

\bibitem{haeusser2017associative}
P.~Haeusser, T.~Frerix, A.~Mordvintsev, and et~al., ``Associative domain adaptation,'' in \emph{Proceedings of the IEEE International Conference on Computer Vision}, 2017, pp. 2765--2773.

\bibitem{xiong2020incipient}
S.~Xiong, Y.~Liu, J.~Fang, J.~Dai, L.~Luo, and X.~Jiang, ``Incipient fault identification in power distribution systems via human-level concept learning,'' \emph{IEEE Transactions on Smart Grid}, vol.~11, no.~6, pp. 5239--5248, 2020.

\bibitem{wu2020harmonic}
J.~Wu, F.~Mei, C.~Chen, and et~al., ``Harmonic detection method in power system based on empirical wavelet transform,'' \emph{Power System Protection and Control}, vol.~48, no.~6, pp. 142--149, 2020.

\bibitem{wang2020power}
W.~Wang, B.~Zhang, W.~Zeng, and et~al., ``Power quality disturbance classification of one-dimensional convolutional neural networks based on feature fusion,'' \emph{Power System Protection and Control}, vol.~48, no.~6, pp. 53--60, 2020.

\bibitem{liu2022method}
Y.~Liu, Y.~Yingjie, S.~Xiong, L.~Pei, Z.~Li, P.~Xu, L.~Su, X.~Fu, and X.~Jiang, ``Method for identifying power equipment targets based on human-level concept learning,'' Mar.~17 2022, uS Patent App. 17/210,530.

\bibitem{li2020fault}
B.~Li, S.~Zheng, H.~Zhan, and et~al., ``Fault diagnosis integrated platform based on typical characteristics of electrical signal on the low voltage side of distribution network,'' \emph{Journal of Electric Power Science and Technology}, vol.~35, no.~6, pp. 92--100, 2020.

\bibitem{xie2016development}
X.~Xie, Y.~Liu, P.~Sun, and et~al., ``Development of novel pmu device for distribution network lines,'' \emph{Automation of Electric Power Systems}, vol.~40, no.~12, pp. 15--20, 2016.

\bibitem{liu2021method}
Y.~Liu, S.~Xiong, C.~Zihan, L.~Lingen, and X.~Jiang, ``Method for detecting power distribution network early failure,'' Oct.~12 2021, uS Patent 11,143,686.

\bibitem{liu2020basic}
Y.~Liu, Z.~Cong, Y.~Yan \emph{et~al.}, ``Basic principles key technologies and development trends of incipient fault detection for power distribution equipments [j],'' \emph{Distribution \& Utilization}, vol.~37, no.~04, pp. 10--16+, 2020.

\bibitem{zhang2020research}
L.~Zhang, Y.~Wang, J.~Liu, and et~al., ``Research and experiments on the transmission characteristics of traveling wave sensor for power fault,'' \emph{Journal of Electric Power Science and Technology}, vol.~35, no.~2, pp. 144--151, 2020.

\bibitem{zhang2019power}
B.~Zhang, J.~Xiao, X.~Liang, and et~al., ``Power outage event completion method based on svm for mv distribution network,'' \emph{Electric Power Engineering Technology}, vol.~38, no.~3, pp. 34--40, 2019.

\bibitem{krizhevsky2012imagenet}
A.~Krizhevsky, I.~Sutskever, and G.~E. Hinton, ``Imagenet classification with deep convolutional neural networks,'' in \emph{Advances in neural information processing systems}, vol.~25, 2012, pp. 1097--1105.

\end{thebibliography}

\end{document}